# 1.4-mJ High Energy Terahertz Radiation from Lithium Niobates

Baolong Zhang[1,3†], Zhenzhe Ma[2†], Jinglong Ma[1], Xiaojun Wu[2,8*], Chen Ouyang[1,3], Deyin Kong[2], Tianshu Hong[2], Xuan Wang[1,4], Peidi Yang[2], Liming Chen[5,6], Yutong Li[1,3,4,6,7*]

[1]Beijing National Laboratory for Condensed Matter Physics, Institute of Physics, Chinese Academy of Sciences, 100190 Beijing, China.

[2]School of Electronic and Information Engineering, Beihang University, 100191 Beijing, China.

[3]School of Physical Sciences, University of Chinese Academy of Sciences, 100049 Beijing, China.

[4]Songshan Lake Materials Laboratory, 523808 Dongguan, Guangdong, China.

[5]Key Laboratory for Laser Plasmas (MOE), School of Physics and Astronomy, Shanghai Jiao Tong University, 200240 Shanghai, China.

[6]Collaborative Innovation Center of IFSA, Shanghai Jiao Tong University, 200240 Shanghai, China.

[7] CAS Center for Excellence in Ultra-intense Laser Science, Shanghai 201800, China.

[8]Huazhong University of Science and Technology, Wuhan National Laboratory for Optoelectronics, Wuhan 430074, China

# Abstract

Free-space super-strong terahertz (THz) electromagnetic fields offer multifaceted capabilities for reaching extreme nonlinear THz optics, accelerating and manipulating charged particles, and realizing other fascinating applications. However, the lack of powerful solid-state THz sources with single pulse energy >1 mJ is impeding the proliferation of extreme THz applications. The fundamental challenge lies in hard to achieve high efficiency due to high intensity pumping caused crystal damage, linear absorption and nonlinear distortion induced short effective interaction length, and so on. Here, through cryogenically cooling the crystals, delicately tailoring the pump laser spectra, chirping the pump pulses, and magnifying the laser energies, we first successfully realized the generation of 1.4-mJ THz pulses lithium niobates under the excitation of 214-mJ femtosecond laser pulses via tilted pulse front technique. The 800 nm-to-THz energy conversion efficiency reached 0.7%, and a free-space THz peak electric and magnetic fields reached 6.3 MV/cm and 2.1 Tesla. Our numerical simulations based on a frequency-domain second-order nonlinear wave equation under slowly varying envelope approximation reproduced the experimental optimization processes. To show the capability of this super-strong THz source, nonlinear absorption due to field-induced intervalley scattering effect in high conductive silicon induced by strong THz electric field was demonstrated. Such a high energy THz source with a relatively low peak frequency is very appropriate not only for electron acceleration towards table-top X-ray sources but also for extreme THz science and nonlinear applications.

\*E-mail: xiaojunwu@buaa.edu.cn,

\*E-mail: ytli@iphy.ac.cn

# Introduction

The special position of terahertz (THz) waves in the electromagnetic spectrum determines its unique physical properties and promising applications, such as low single-photon energy, vibrational fingerprint of macromolecules in (bio)materials, high-speed line-of-sight wireless communication capabilities, and so on[1-4]. During the past several decades, THz science and technologies have experienced rapid progress and start to move from the laboratory research stage to practical applications[5-7]. However, one of the key factors hindering the THz development lies in the severe lack of strong THz sources. Along with the fast development of ultrafast laser technologies, pulsed THz sources and other related systems pumped by femtosecond lasers have become ubiquitous[8-16]. Nevertheless, many THz science and applications are primarily focused on weak-field passive detection of matter, based on linear light-matter interaction process. On the other hand, intense THz field-induced nonlinearly active control of matter[17-36] is, to some extent, still out of reach due to the shortage of highly efficient, and stable THz sources. It results in many mesoscale/microscopic physics and phenomena almost inaccessible. Although ultrafast microjoule-scale THz sources have already exhibited their powerful capabilities in electron acceleration and manipulation[37-42], field-induced phase transition[17-20], and nonlinear THz phenomena[24-31] in a wide range of materials and structures, the demand for millijoule-level THz sources has not yet been fully met, especially when facing the dawn of extreme THz science overwhelmingly growing to the next frontier in nonlinear optics[29, 43].

Historically, the first THz pulse was produced in a lithium niobate (LN) wafer driven by a pulsed laser via optical rectification[44]. However, due to difficult phase matching and large linear absorption coefficients[45], LN crystals have been neglected by THz community for more than thirty years, while other nonlinear crystals such as ZnTe, GaP, and GaSe are more favored[46]. In 2002, the tilted pulse front (TPF) technique was proposed and LN crystals came back to the THz research field[47]. During the recent twenty years, LNs have illustrated its outstanding potential for efficient THz generation, accompanied by a large boosting of THz single pulse energy from femto-joule to microjoule and recently to sub-millijoule[48]. In parallel, other methods for generating strong field THz radiation, such as organic crystal-based optical rectification[49], laser-solid interaction[11], two-color air plasma scheme[8, 9], laser pumped liquid-plasma[15], and so on, are also experienced unprecedented prosperity. For air-plasma sources, the conventional two-color fields induced air plasma has made great progress recently and has successfully obtained 0.19 mJ THz pulses with the conversion efficiency of 2.36% when employing mid-infrared high energy laser pulses at 3.9 μm as the pump laser[9]. Plasma in liquids can also generate THz radiation, and a 0.07-mJ pulse energy source with the conversion efficiency of 0.27%[15] was demonstrated. The highest single pulse THz energy of tens of mJ was achieved from solid-plasma pumped by 60 J laser pulses via transition radiation mechanism[11]. Although these strong sources make it possible for the stepping forward towards extreme nonlinear THz investigations, nonlinear crystal-based solid-state intense THz sources such as organic crystals and LNs are more widely utilized due to their high optical-to-THz conversion efficiency. From organic crystal-based THz sources, 0.9-mJ single pulse THz energy has been demonstrated[49] and such sources were used for discovering many resonant and nonresonant nonlinear THz effects[29]. LNs assisted by the TPF technique are also favored not only due to its high efficiency and high damage threshold but also featuring intrinsic high stability and reliability for applications.

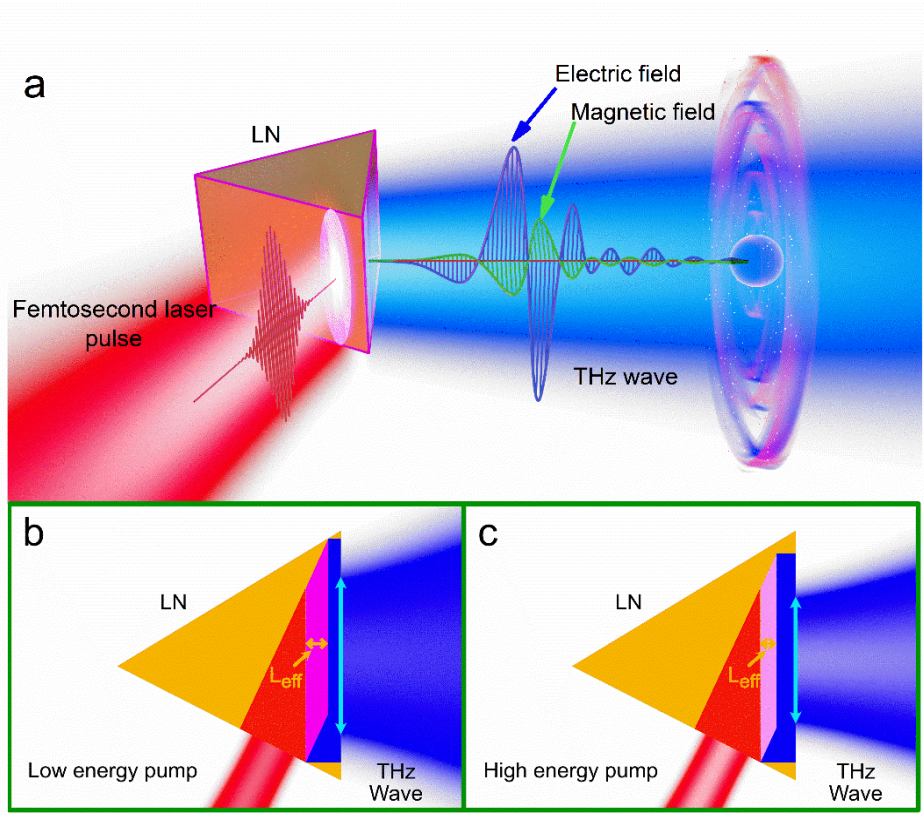

**Fig. 1 Schematic diagram of high energy THz generation from LNs via TPF technique. a,** High power ultrashort femtosecond laser pulses illuminate onto LN crystals can produce appreciably high energy THz pulses for extreme THz science and nonlinear investigations. Comparison of the effective interaction lengths under **b,** moderate and **c,** higher power pumping due to nonlinear distortion effect. Higher pumping fluence corresponds to shorter effective interaction lengths, resulting in saturation or reduction of the THz yields.

However, the THz energy record from LNs has been held at 0.4 mJ for a long time[48], much lower than that from the piece-by-piece joint organic crystals[49]. To further boost THz energies up to 1 mJ, one may intuitively think that the simplest way is to scale up the pump laser energy together with enlarging the LN crystal size (see Fig. 1a). However, under such circumstances, effective interaction lengths inside crystals for phase matching will become tremendously short due to nonlinear distortion effect in the non-collinear geometry scheme in TPF setup[50], as schematically illustrated in Fig. 1b and c. When the phase-matching condition is successfully satisfied as soon as the pump laser pulses entering the LN incidence facet, high optical-to-THz conversion efficiency will be achieved, resulting in great deterioration for the pumping spectrum. In the subsequent propagation of the pumping laser pulses, the ability to generate THz radiation will be strongly weakened. Consequently, according to the optical-to-THz conversion efficiency formula (1), the conversion efficiency will be extremely difficult to be maintained due to reduced effective interaction lengths.

$$\eta_{THz} = \left( \frac{2d_{eff}^2}{\varepsilon_0 n_{opt}^2 n_{THz} c^3} \right) \Omega_{THz}^2 \cdot I \cdot \left[ L^2 \exp\left( \frac{-\alpha L}{2} \right) \cdot \frac{\sinh^2(\alpha L/4)}{(\alpha L/4)^2} \right] \quad (1)$$

where $\eta_{THz}$ denotes THz efficiency; $d_{eff}$ refers to the effective nonlinear coefficient; $\varepsilon_0$ is the vacuum permittivity; $n_{opt}$ is 2.3 corresponding to the optical refractive index of LN at 800 nm, while $n_{THz}$ is 4.9 for LN at 1 THz; $c$ is the speed of light in vacuum; $\Omega_{THz}$ is the THz angular frequency; $I$ is the pump intensity; $L$ is the effective interaction length; $\alpha$ represents the THz absorption coefficient. What makes things even worse is that commercialized high power femtosecond lasers in most laboratories or factories are mostly based on Ti:sapphire laser technologies. These laser systems can deliver 800 nm central wavelength with tremendously abundant pump energies for THz generation[51,52]. However, such laser technologies are good at producing ultrashort laser pulses normally ≤50 fs pulse duration, which intrinsically further limit effective interaction length in the optical rectification process[53]. Although in formula (1), THz efficiency could be enhanced by increasing pumping intensity due to ultrashort pulse widths, crystal damage and multiphoton absorption effects hinder the usage of ultrashort pulse induced high pump intensity, resulting in reduced THz efficiency. All these aforementioned physical and technical challenges summed up together make 1 mJ pulse energy a very difficult barrier from LNs pumped by ultrashort ultra-strong Ti:sapphire laser pulses. Therefore, low pump energy laser pulses illuminate LNs are used for producing THz pulses, to some extent, with relatively high conversion efficiency, but suffering from low absolute THz energy. While LNs are illuminated by high power ultrashort ultra-strong laser pulses, we are chasing for the extremely high single pulse THz energy with comparable conversion efficiencies, since real applications concern about absolute THz energies (photon numbers) and free-space peak field strengths interacting on the matter.

In this work, we successfully overcome the aforementioned challenges in the process of producing >1 mJ THz pulses from LNs via the TPF technique and obtained a 1.4 mJ record number for this method. Through delicately regulating the pump spectra, shaping and enlarging the pump laser spot, chirping the laser pulses, and cryogenically cooling LN crystals, we not only achieved a new record number for 800 nm-to-THz conversion efficiency of ~0.7% under 214 mJ pump laser energy but also realized monotonously enhancement of the radiated absolute THz energy during this generation process. The emitted super-strong THz pulse bullets are comprehensively diagnosed by single-shot interferometry technique, and its peak frequency was located at 0.4 THz with a 6.3 MV/cm peak electric field and 2.1 Tesla magnetic field strength, respectively. Numerical simulations can well interpret our experimental results, and a strong field THz nonlinear verification experiment was implemented in a doped silicon wafer with high conductivity. We believe our work is helpful for further understanding intense THz radiation physics, and the super-strong THz sources in the low-frequency range are ready for advanced nonlinear THz investigations and other fascinating application experiments.

## Experimental implementation

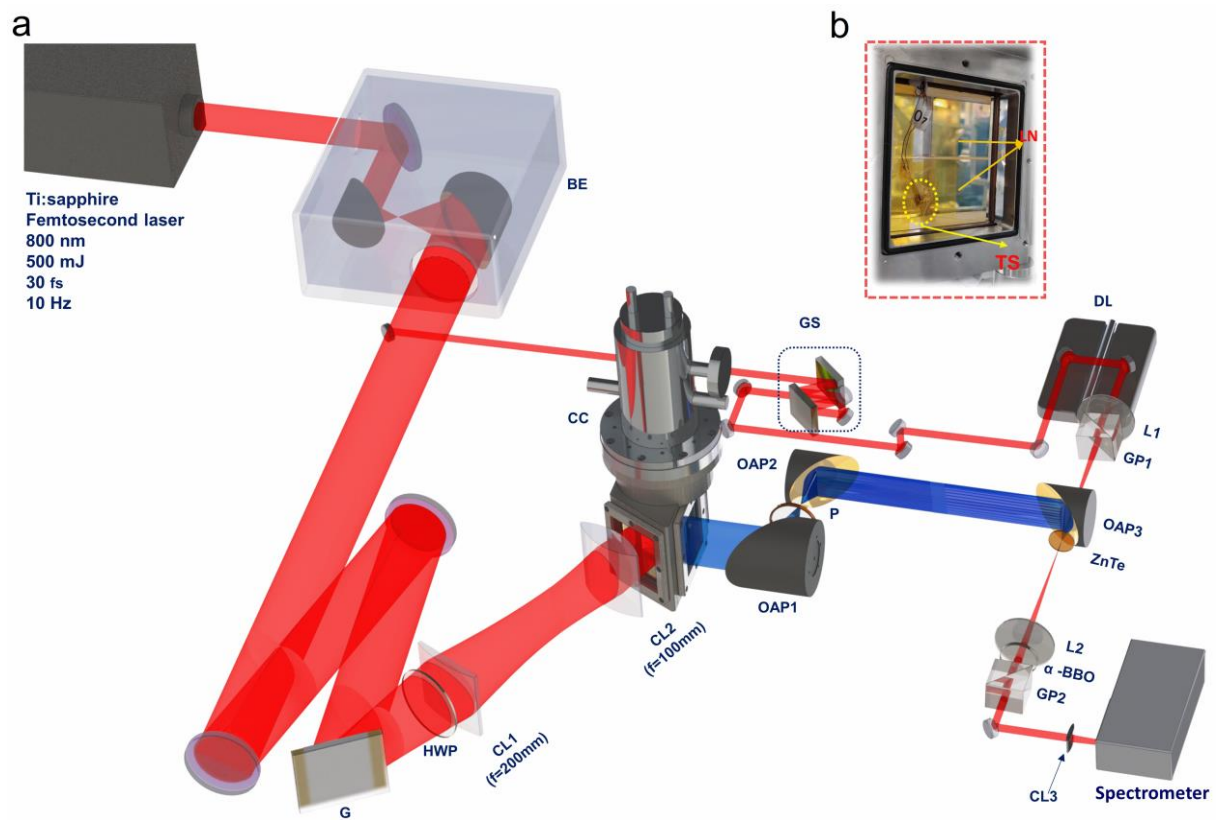

**Fig. 2 Experiment setup for THz generation and characterization. a,** The pump pulses from a commercial Ti:sapphire laser amplifier was first expanded and then guided to a grating (G). The diffracted pump beam first passed through a half wave plate (HWP), and then two cylindrical lenses (CL1 and CL2), and was incident onto the LN crystals for THz generation. The crystals were cryogenically cooled in a cooling chamber (CC) by liquid nitrogen. The outcoupled THz pulses were first collimated and guided by two parabolic mirrors (OAP1 and OPA2), and then focused by another parabolic mirror (OAP3) with a hole (diameter of 4 mm) together with the stretched probing beam sampled from the pump beam just after the beam expander. For single-short diagnostic measurement, the probing beam passed through a delay stage (DL1), a focus lens (L1), and a Glan prism (LP1). After the detection crystal ZnTe, the probing beam was collimated by another lens (L2), and illuminated onto a BBO crystal and a Glan prism (GP2) for entering the spectrometer. CL3: Cylindrical lens. **b,** The photo of LN crystals mounted in the cooling chamber. TS: temperature sensor.

In our experiment, as shown in Fig. 2a, we employed a commercial Ta:sapphire femtosecond laser (Pulsar20, Amplitude Technology) being able to deliver the maximum pump energy of 500 mJ with center wavelength 800 nm, pulse duration 30 fs, and repetition rate 10 Hz[52]. The pump laser spectrum and group dispersion delay can be adjusted by a device Dazzler (Acousto-optic programmable dispersive filter, Fastlite) equipped in this laser system. To prevent damaging the LN crystals, the pump laser was expanded to a beam diameter of 80 mm ($1/e^2$) through a beam expander consisting of two 90° off-axis parabolic mirrors placing in a vacuum chamber to avoid air breakdown. The expanded pump beam was guided by two reflection mirrors and incident onto a grating (1500 grooves/mm). For the phase-matching requirement, the incidence angle was 21°, resulting in minus one maximum diffraction at 57°. The diffracted beam passed through a 4-inch half-wave plate and its polarization was varied from p-polarized to s-. The 4f imaging scheme consists of a pair of confocal cylindrical lenses with their focal lengths of 200 mm and 100 mm resulting in a demagnification factor of -0.5. The pump laser beam spot on the LN crystal was elliptic with a long axis (vertical) of ~80 mm and a short axis (horizontal) of ~10 mm.

The two congruent LN crystals (z-cut, 68.1×68.1×64 mm$^3$ in x-y plane, height~40 mm) for THz generation were placed at the focal plane of the second lens[52]. This kind of configuration can tilt the pulse front angle to 63° satisfying the phase-matching condition between the generated THz wave and the infrared pulses in the LNs and reducing the output THz divergence[50,54]. Since the commercialized LNs with 80 mm height in the z-axis are not available, we used two LN prisms and stacked them in the z-axis, as exhibited in Fig. 2b. An indium foil layer was tightly attached between them to keep their temperature consistent. To reduce the THz linear absorption and enhance the effective interaction length, the crystal was cryogenically cooled by liquid nitrogen. The crystal temperature was monitored during the experiments. To minimize the Fresnel loss at the THz output facet, an anti-reflection layer with ~96 μm thickness including three stacked Kapton tap layers (polyimide, Thorlabs, KAP22-075) was tightly attached on the THz output surface of LNs[51, 52].

High energy THz pulse diagnostic included energy and efficiency measurement and single-shot temporal waveform characterization. The former was conducted by a calibrated pyroelectric detector (Gentec SDX-1152) placing at the focus of the first parabolic mirror after LN crystals together with many THz attenuators (see Supplementary Information). The focused THz beam profile was measured by a commercial THz camera (Ophir, Pyrocam IV). The single-shot interferometry measurement included a ZnTe detector, a BBO crystal, a Glan-prism, a stretcher with a pair of gratings, and a spectrometer. More details can be found in methods.

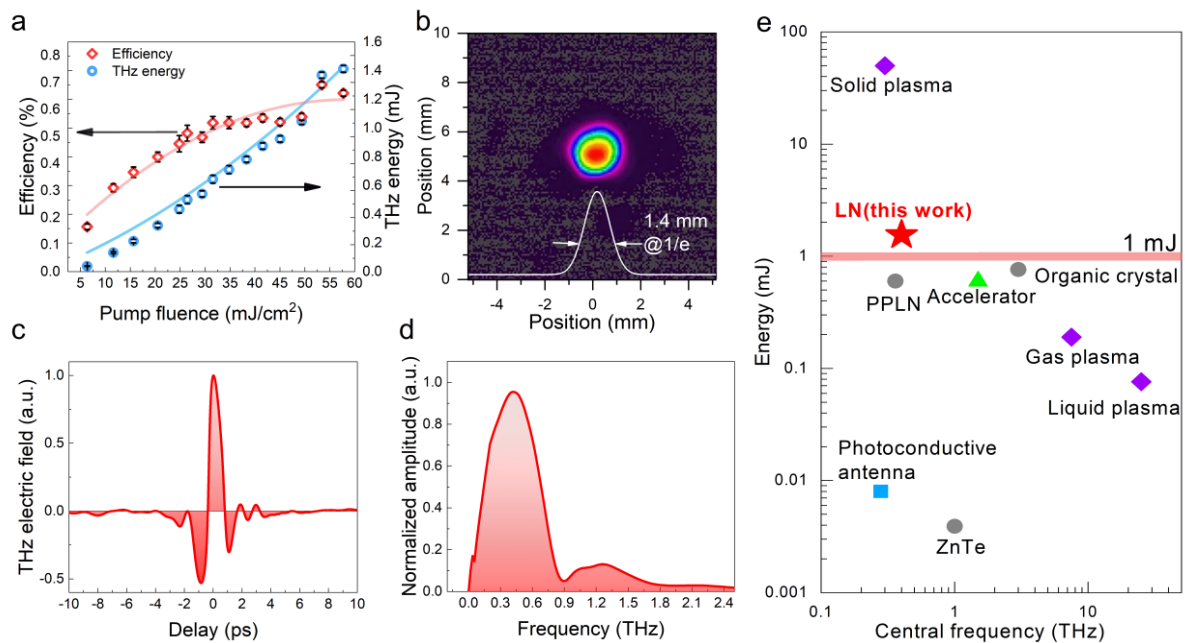

**Fig. 3 Characterization of the experimentally realized high energy THz pulses. a,** The obtained new record number of a 1.4 mJ high energy THz pulse from LN crystals, and the extracted THz energy (blue dots) as well as its corresponding conversion efficiency (red dots) as a function of the pump fluence. **b,** Focused THz beam profile with a diameter of ~1.4 mm (1/e) measured at the focus of OAP1. **c,** A typical THz time-domain waveform with 1 ps (FWHM) pulse duration probed a single-shot measurement, and **d,** its corresponding spectrum with a peak frequency at 0.4 THz. **e,** Comparison of current record THz sources. These data

are referenced from previous record results of THz sources based on solid plasma[11], liquid plasma[15], gas plasma[9], optical rectification from ZnTe[60] and organic crystals[49], different frequency generation in PPLNs[13], accelerators[61], and photoconductive antennas[62].

The experimentally achieved highest THz pulse energy was 1.4 mJ, as plotted in Fig. 3a, with a maximum 800 nm-to-THz conversion efficiency is ~0.7% under the infrared pump energy of 214 mJ, when the LN temperature was 93 K. To obtain this record number in LN-TPF method, the group delay dispersion and spectral width of the pump beam were set as 2200 fs$^2$ and 11 nm (FWHM), respectively. Besides, fine-tuning of the TPF setup was extremely significant, and the pump spectral distribution should also be delicately engineered. After these optimizations, we measured the pump fluence dependent THz energy and its corresponding efficiency curves (Fig. 3a). As we can see, the recorded THz energy obeys a monotonous boost tendency along with the increase of pump fluence without showing any saturation sign. For the efficiency curve, it linearly increases till to 58 mJ/cm$^2$, and then starts to saturate slowly. However, even we applied the maximum pump energy of 214 mJ, the efficiency did not show reduction indication. For practical nonlinear applications, absolute THz energies play the key role, implying that higher THz energies could be continuing boosted by scaling up the pump energy to the tolerance of the crystal damage threshold (estimated value of 100 mJ/cm$^2$).

Fig. 3b illustrates the measured THz beam profile. The focused beam has a diameter of 1.4 mm (1/e). The ideal Gaussian beam exhibits a nearly flawless profile, phenomenologically manifesting the feasibility of stacking two LN crystals as an integrated large-size emitter. The probed typical THz temporal waveform and its corresponding spectra by the single-shot method are exhibited in Fig. 3c and Fig. 3d, respectively. Its pulse width is estimated to be ~1 ps (FWHM). The emitted peak frequency is located at 0.4 THz, and its high-frequency component can extend to 2 THz, which is very appropriate for THz electron acceleration and other field-controlled light-matter interaction investigations requiring low photon energies and ultrafast time-resolution. With these parameters of single pulse energy, pulse duration, and beam size, the calculated focused THz peak electric field is 6.3 MV/cm with a pulsed magnetic field strength of 2.1 Tesla. Such free-space THz field strengths can be further improved by sufficiently tightly focusing the beam through delicately designing focusing optics down to its diffraction limit[55], and the ideally focused electric field strength can be larger than 10 MV/cm. More details about the calculation can be found in Supplementary Information.

For the THz energy of 1.4 mJ with its peak frequency at 0.4 THz, the total photon number ($N$) contained in a single pulse is ~$5.4 \times 10^{18}$ calculated by using $N = E_T / (h \times \upsilon_P)$, where $E_T$, $h$ and $\upsilon_P$ denote the energy of THz pulse, Planck's constant, and the peak frequency of THz pulse, respectively. Such a high photon number is compared to that of an 800 nm laser pulse with a single pulse energy of 1.35 J. We summarized the current high energy THz sources with record numbers in Fig. 3e. There are only two kinds of intense THz sources with single pulse energies larger than 1 mJ. The 1.4 mJ THz energy is the first high energy solid-state THz source with single pulse energy higher than 1 mJ, which is 1.6 times of organic crystal record number[49], 3.5 times of that from current LN-TPF driven by 1 μm central wavelength[48], 7.4 times that from both two-color air plasma sources[9] and LN wafers driven by 2 J Ti:sapphire lasers[56]. Although the absolute THz energy in our case is not that high compared with solid-plasma THz sources via transition radiation mechanism[11], the solid-plasma sources require tens of joules pump laser energies which are not easy to be accessible. LN-based intense THz sources are much easier to implement and less expensive. Furthermore, our demonstrated 800 nm-to-THz energy conversion efficiency is also a new record number in LN-based sources. Although 1 μm pumped LN-based THz sources have created a conversion efficiency of 0.77%, it is more challenging to realize a comparable efficiency under the excitation of 800 nm laser pulses due to easier multiphoton generation in LNs with 4.0 eV bandgap[51]. Besides, the polarization purity, divergence, and directionality of the LN-based source are much better than transition radiation sources, ensuring lower energy loss during the propagation process and guaranteeing higher absolute energy available at samples in subsequent application experiments.

During the realization of the super-strong THz radiation process, the LN crystal temperature is also essential for phase matching, effective interaction length, and THz linear absorption. Phase matching is influenced by the refractive index of LNs, and under different temperatures, the refractive index is different[45], leading to the requirement of fine tuning of the TPF setup at different crystal temperatures. Furthermore, microscopically, at a low crystal temperature, THz photons inside crystals are less absorbed by lattice vibrations, coherently boosting the macroscopic effective interaction length. Thirdly, THz output energies are severely affected by LN crystal linear absorption due to lattice vibrations. There are six phonon absorption located at 3.9 THz, 7.4 THz, 8.2 THz, 5 THz, 18.8 THz, and 20.8 THz[57]. During the generation process, the first two-photon vibrations dominant the primary THz linear absorption.

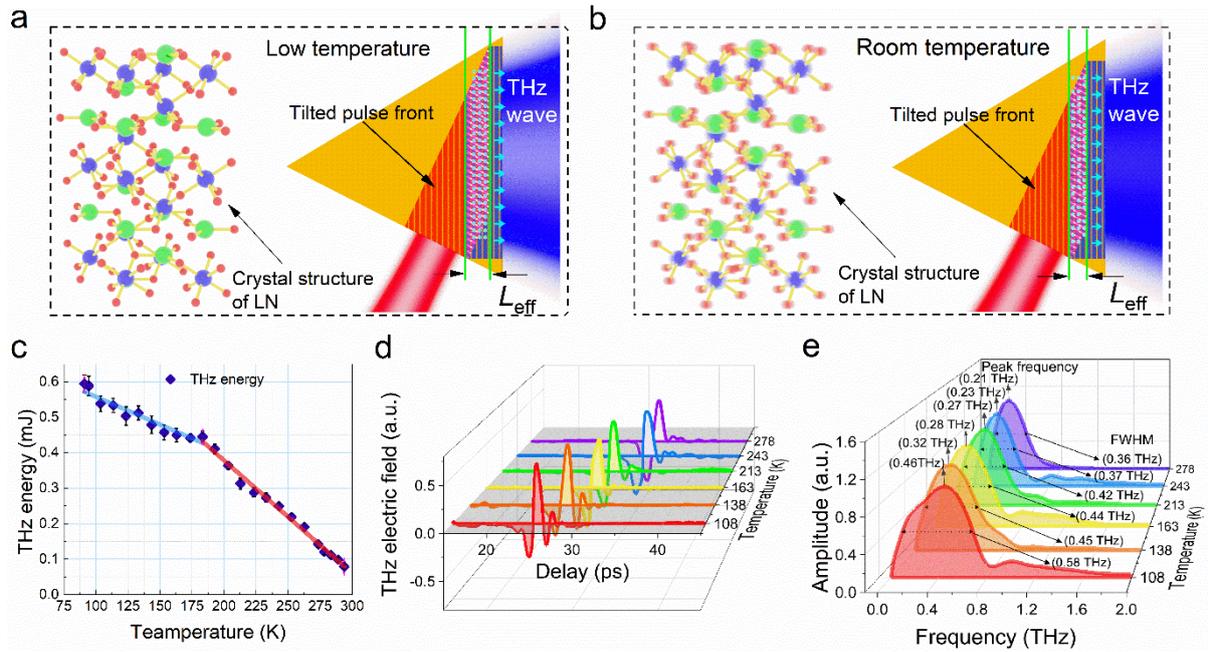

**Fig. 4 The influences of LN crystal temperature on THz yields. a,** Schematic diagram of the enhanced effective interaction lengths and reduced THz absorption in the THz generation process from lower temperature LNs by the cooling method and **b,** room temperature case. **c,** The generated THz energy as a function of the LN crystal temperature. **d,** Typical THz time-domain waveforms, and **e,** their corresponding spectra.

To efficiently extracting the generated THz energy as well as obtaining good phase matching, we investigated the collected THz energies and their corresponding waveform transients as a function of the LN crystal temperature. For this measurement, we optimized the TPF system at 91 K with the pump energy of 180 mJ. Then we waited for the crystal temperature back to 300 K, and the data was recorded when the crystal was once again cryogenically cooled down to the low temperature. As we can see in Fig. 4a, the THz energy keeps monotonous increasing trend during the whole cooling process, and the energy at 91 K is 6 times higher than that at room temperature. However, it is worth noting that the energy enhancement rate from 300 K to 183 K is more precipitous than that from 184 K to 91 K, implying a potential weak saturation effect for the crystal temperature on THz output energy[58]. Although the temperature-dependent THz yield curves may exhibit slightly different line shapes for different optimization conditions, the general tendency is almost the same, indicating cryogenically cooling is a very effective way for achieving extremely high THz energies.

To further clarify what kind of THz photons are easier to escape from lattice vibration trapping, we systematically measured the output THz temporal waveforms dependent on the crystal temperature via single-shot methods. As depicted in Fig. 4c, the first observed phenomenon was the remarkably advanced time delay for the radiated THz signals when the crystal temperature was reduced. This experimental behavior further corroborates the aforementioned reduction of the crystal refractive index and identifies the high efficiency at low temperatures. Besides, along with the decreasing of the crystal temperature, the radiated THz pulse width becomes strikingly narrower. Consequently, as depicted in Fig. 4e, the Fourier transferred emission spectra illustrate appreciable broadened behavior and their peak frequencies saliently blueshift, manifesting that cooling technique can decisively assist THz photons with higher energies to propagate inside LN crystals, resulting in efficiently coupled out with lower heat dissipation.

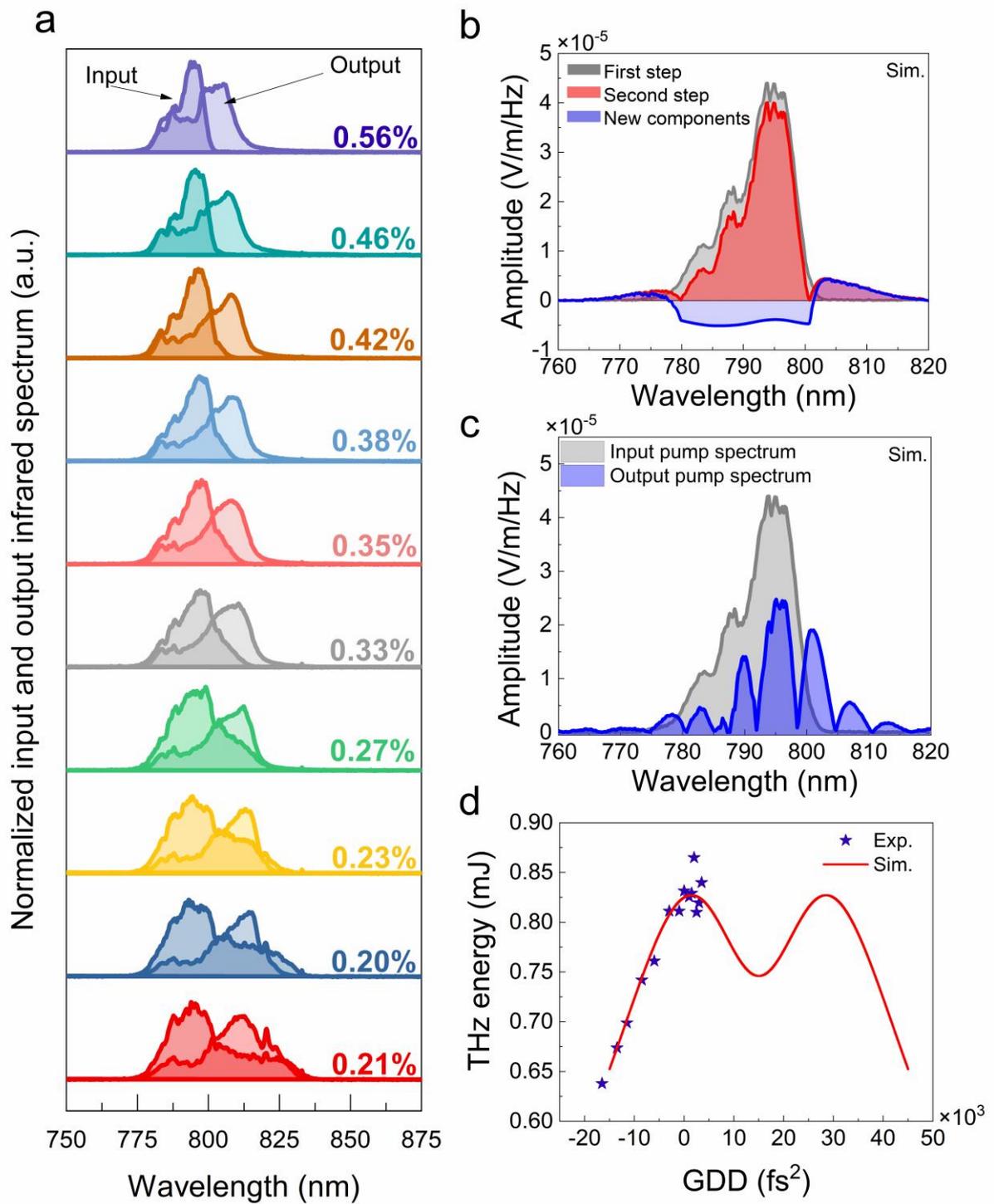

**Fig. 5 Tailoring pump spectrum and chirping the pump pulses on THz efficiencies. a,** Input and output pump laser spectra and their corresponding THz efficiencies. Tailoring the pump spectra can produce higher THz efficiencies. THz radiation process led to the input spectra broadened themselves and central frequencies redshifted. **b,** Simulative laser amplitude

spectra of the first two steps and the new frequency components (the difference between the two spectra) generated in "the second step". **c**, Simulative output laser spectrum after the whole simulation process and the corresponding input laser spectrum (taking from input spectrum with the highest efficiency in **a**). **d**, The experimental data theoretical result of THz energy as a function of group velocity dispersion.

To make things more intriguing, we can gain more useful information about light-matter nonlinear interaction physics and further disentangle the influence factors of the THz yields by comprehensively analyzing the input and output infrared spectra. During the THz generation process, we found a notable effect that elaborately tailoring the pump spectrum can appreciably enhance the THz efficiency, as illustrated in Fig. 5a. When the input pump laser spectra from the Ti:sapphire laser amplifier were broadband distribution with relatively symmetric shapes, the generation efficiency was considerably small. Surprisingly, when we engineered the pump laser spectra by operating the built-in acousto-optic programmable dispersive filter (AOPDF, Dazzler, Fastlite), the extracted THz efficiency was significantly enhanced. For this investigation, we fixed the pump energy at 180 mJ and tailored the pump spectrum, which promoted the THz efficiency from 0.21% to 0.56%, resulting in more than twice productiveness.

When considering the corresponding output laser spectra, as plotted in Fig. 5a, it is obvious that the output spectral widths were tremendously broadened, and an obvious redshift phenomenon was also observed for all efficiencies when compared with the input spectra (Fig. 5a). Based on the input and output infrared spectra, we can safely estimate the intrinsic optical-

to-THz conversion efficiency according to the following equation

$$\eta = 1 - \lambda_i / \lambda_o \qquad (2)$$

where $\lambda_i$ is the center wavelength of the input spectrum, $\lambda_o$ denotes the center wavelength of the output wavelength. With this method, the calculated estimated intrinsic efficiency for the experimentally extracted efficiency of 0.56% can be as high as 1%, for which the input and output central wavelengths are 794 nm and 802 nm, respectively. The losses may come from (a) residual Fresnel loss for both pump laser and THz waves at the incidence and output facets, respectively, (b) remaining crystal absorption for THz photons due to non-thoroughly eliminating lattice vibrations by cryogenically cooling, and (c) transmission and collection losses.

Last but not the least, we also experimentally studied the impact of chirping the pump pulses on THz efficiencies, and the chirping was also realized by tuning the built-in acousto-optic programmable dispersive filter. As summarized in Fig. 5c, both the probed THz energy and its corresponding efficiency reached their peak values when the group delay dispersion was 2200 fs$^2$ corresponding to 110 fs. Combining with our previously published data[52], we can conclude that no matter for moderate or ultra-strong THz energy generation, chirping the pump laser pulses is very helpful to improve the THz productiveness, especially for ultrashort laser pulses.

**Numerical simulations**

To quantitatively describe the complete physical picture of the second-order nonlinear optical propagation of the femtosecond laser pulse in the LNs, we derived a second-order nonlinear frequency-domain wave equation under slowly varying envelope approximation.

$$\frac{\partial E_{env}(\omega,z)}{\partial z} = \frac{-j\mu_0\omega^2}{2k(\omega)} \frac{Ne^3 a}{2\pi m^2 D(\omega)} \int_{-\infty}^{+\infty} \frac{E_{env}(u,z)e^{-\alpha(u)z} E_{env}(\omega-u,z)e^{-\alpha(\omega-u)z}}{D(u)D(\omega-u)} du \cdot e^{\alpha(\omega)z} \quad (3)$$

where $E_{env}(\omega,z)$ denotes the frequency-domain complex envelope of electric fields, $j$ is the unit of imaginary number, $\mu_0$ denotes vacuum permeability, $k(\omega)$ is the wavenumber which can be described by $k(\omega) = \beta(\omega) - j\alpha(\omega) = \omega\sqrt{\varepsilon\mu_0}$ ($\varepsilon$ means permittivity of LN crystals), where $\alpha(\omega)$ presents absorption coefficient, $N$ denotes atomic number density, $e$ is the electron charge, $m$ denotes electron mass, and $a$ presents nonlinear coefficient for nonlinear Lorentz oscillator model. The factor $D(\omega)$ is described by $D(\omega) = -\omega^2 + \omega_0^2 + j\gamma\omega$, where $\omega_0$ is the eigenfrequency and $\gamma$ is damping rate.

This frequency-domain wave equation is a first-order ordinary differential equation with respect to the propagation length $z$, hence it is available to be solved by numerical calculation and simulation. Taking the input laser spectrum before entering the crystal as the boundary condition $E_{env}(\omega, z=0)$, we can employ the fourth-order Runge-Kutta method to numerically solve the complex envelope of the electric field of light $E_{env}(\omega,z)$ for arbitrary propagation distance $z$, while the actual light electric field can be described as $E(\omega,z) = E_{env}(\omega,z)e^{-jk(\omega)z}$.

We used the full-band bilateral spectrum in this wave equation and treated electromagnetic waves of arbitrary frequencies as an integral without distinguishing the THz electric field and the optical electric field. All frequency components of light were concentrated in the electric field complex envelope $E_{env}(\omega,z)$. With this method, we managed to characterize the continuous variation of $E_{env}(\omega, z)$ with respect to the propagation length $z$ precisely using equation (3).

To interpret the various phenomena of the pump laser spectrum, we considered the laser pulse entering the LNs and propagating a tiny distance $dz$, defined as "the first step". The newly produced frequency components were approximately described by $\left.\frac{\partial E_{env}(\omega, z)}{\partial z}\right|_{z=0} \cdot dz$ which were generated by the convolution term on the right side of equation (3). Assuming that the initial laser spectrum distributed between $\omega_{min}$ and $\omega_{max}$, in other words, $\omega_{min} \leq |\omega| \leq \omega_{max}$, due to mathematical properties of convolution, the new frequency components caused by the convolution term were distributed into two regions. The first region of $0 \leq |\omega| \leq \omega_{max} - \omega_{min}$ represents the difference-frequency generation (DFG) and optical rectification, while the second region of $2\omega_{min} \leq |\omega| \leq 2\omega_{max}$ describes sum-frequency generation (SFG) and second-harmonic generation. The former is exactly in the THz frequency range. Therefore, all second-order nonlinear optical processes are contained in the convolution term. The new frequency components in "the first step" have a 90° phase lag comparing to the initial laser spectrum due to the existence of $-j$.

When all the electromagnetic waves propagate again a tiny distance $dz$, defined as "the second step", the former DFG components (THz frequencies) generated from "the first step" which was located in $0 \leq |\omega| \leq \omega_{max} - \omega_{min}$ continue to make convolution with the laser spectrum located in $\omega_{min} \leq |\omega| \leq \omega_{max}$, generating frequency components in $\omega_{min} - (\omega_{max} - \omega_{min}) \leq |\omega| \leq \omega_{max} + (\omega_{max} - \omega_{min})$. These components in "the second step" represent the DFG and SFG processes between the THz radiation and laser pulse, and have a

180° phase lag comparing to the initial laser spectrum. The laser spectrum then broadened itself to the low-frequency region in $\omega_{min} - (\omega_{max} - \omega_{min}) \leq |\omega| \leq \omega_{min}$, and the high-frequency region in $\omega_{max} \leq |\omega| \leq \omega_{max} + (\omega_{max} - \omega_{min})$, but depleted the initial spectrum due to the phase difference of 180°. The simulation results of the first two steps are plotted in Fig. 5b. The amplitude of the new frequency components generated in "the second step" (blue line) is negative in the initial laser spectrum region but is positive outside From these numerical results, it is clearly shown that during the two steps of generation, the input laser spectrum amplitude was reduced accompanying the production of new frequency components at its shoulders on both sides. This reproduces the spectrum broadening phenomenon observed in our experiment (Fig. 5a). As all the electromagnetic waves continue to propagate in the LN crystal, the laser pulse and THz radiation exchange energy to each order according to the convolution and phase differences, leading to a broadening and depletion of the laser spectrum (Fig. 5c). Hence the overall trend is that parts of the laser energy covert to THz energy.

To explain the physical picture of tailoring the input pump laser spectrum on the influence of THz efficiency, assuming that the laser spectrum is asymmetric with respect to the peak frequency and contains more frequency components on the high-frequency side than that in the low-frequency region, the DFG and SFG processes between the THz radiation and laser pulse mainly perform as depletion on the high-frequency side but mainly perform as broadening on the low-frequency side. This asymmetric phenomenon makes the laser spectrum obtain much more low-frequency (long-wavelength) components than the other side, resulting in the laser spectrum moving to the low-frequency (long-wavelength) side. This phenomenon is verified by the blue curve in Fig. 5b, which new frequency components on the low-frequency (long-wavelength) side are more. Continuing to simulate more steps, we obtained the final output laser spectrum and plotted together with the initial laser spectrum in Fig. 5c. The input laser spectrum used in the simulation is the spectrum with an efficiency of 0.56% in Fig. 5a. Remarkable depletion, broadening, and redshift can be observed in Fig. 5c after the nonlinear propagation, which is consistent with the experimental tendency. The frequency redshift behavior of the laser spectrum makes the laser energy flow to THz frequency range, and the asymmetry of the initial laser spectrum pushes enables more laser energies to convert into THz energies, consequently increasing the optical-to-THz energy conversion efficiency. This mechanism provides a good explanation of our experimental results that asymmetric input spectrum with more high-frequency components produces enhanced THz yields, the broadened and the red-shifted output infrared spectrum.

The essence of adding linear chirp on an input laser pulse is employing a linear second-order dispersion phase term $\frac{\beta}{4}\omega^2$ on the frequency-domain complex envelope of the laser pulse, where $\beta$ is the chirp parameter. To theoretically describe the variation of conversion efficiency as a function the group velocity dispersion, two assumptions are imposed in this case: (a) the laser possesses a Fourier transform limited Gaussian spectrum which remains unchanged during the whole propagation process; (b) the effective second-order susceptibility $\chi_{eff}^{(2)}$ is regarded as a frequency-independent constant. Therefore, the frequency-domain complex envelope of the laser pulse can be expressed as

$$A_{op}(\omega) = \frac{E_0 \tau}{2\sqrt{\pi}} e^{-\frac{(\tau^2 + j\beta)\omega^2}{4}} \quad (4)$$

where $\tau$ is the Fourier transform limited half pulse width ($e^{-1}$), $\beta$ is the group velocity dispersion constant, and $E_0$ describes the intensity of the laser pulse. The generated THz electric field spectrum under this circumstance is expressed as follows[59],

$$E_{THz}(\Omega, z) = -j\frac{\Omega^2 \chi_{eff}^{(2)} E_0^2 \tau_{eff} e^{-\frac{\Omega^2 \tau_{eff}^2}{8}}}{c^2 2\sqrt{2\pi}} \frac{e^{-jk(\Omega)z}}{\alpha_{THz}(\Omega)k(\Omega)} \cdot (1 - e^{-\frac{\alpha_{THz}(\Omega)z}{2}}) \quad (5)$$

where $\chi_{eff}^{(2)}$ is the effective second-order susceptibility, $\alpha_{THz}(\Omega)$ is the absorption coefficient in the THz range, $k(\Omega)$ is the wavenumber in THz range. $\tau_{eff}$ is defined as "the effective pulse width", which is the most essential variable in this expression. It represents the equivalent Fourier transform limited pulse width $\tau$ which generates the same THz electric field spectrum as the linear chirped laser. The expression of $\tau_{eff}$ is defined as

$$\tau_{eff} \equiv \sqrt{\frac{\tau^4 + \beta^2}{\tau^2}} \quad (6)$$

The linear chirp induces an effective broadening for the pulse width regardless of the sign of $\beta$, consequently broadening the THz electric field spectrum, while maintaining the amplitude of the laser spectrum unchanged. However, due to the strong vibrational absorption to high-frequency components in the THz region, the amplitude of the THz electric field spectrum decreases when its width increases. Hence there is an optimal effective pulse width for the maximum generation of THz energy. Since the initial laser pulse was not dispersionless and it has already carried anomalous dispersion (negative chirp) before artificially adding chirp, there was some offset during the fitting process, leading to a translation of the symmetric simulation curve. We applied the numerical simulation results to fit the experimental data in Fig. 5d, and the qualitative match was good. Based on these results, it was predicted that further chirping the pump pulses would decrease the THz efficiency to the minimum, but it would later recover and reduce again following the same tendency in our previous results[52]. Such a phenomenon has also been reported in high energy THz generation from LN wafers driven by 2 J laser pulses[56].

**High energy THz application**

To further demonstrate the powerful capability of our high energy strong-field THz source on nonlinear investigations, we also performed a nonlinear absorption experiment on both doped and high resistivity silicon wafers. Given that the polarization of the generated THz pulses was vertically polarized, we adjusted the THz pulse energy as well as its electric field strength by a THz polarizer (Tydex). It was placed in front of the focus of OAP1 (see Fig. 2). The 0° shown in Fig. 6 means the maximum THz transmission. The pyroelectric detector was placed right behind the sample to record the transmitted THz energy after samples. The maximum THz pulse energy in this experiment was 0.37 mJ resulting in a focused peak electric field strength of ~3.2 MV/cm. As depicted in Fig. 6, a considerable nonlinear absorption phenomenon was observed in the 0.5 mm-thick doped silicon wafer (doping density ~$10^{16}$ $cm^{-3}$). However, the THz field-induced nonlinear behavior was not be probed in the high-resistivity silicon (~8 kΩ·cm). This nonlinear absorption phenomenon in the doped silicon wafer can be well explained by the intervalley scattering effect of the electrons in the conduction band of the doped silicon[15], as schematically illustrated in Fig. 6b. Based on the microscopic physical picture, the electrons in the low energy valleys can be excited by strong THz fields to high energy valleys in the conduction band. Electrons occupying in the high energy valleys have larger effective mass, resulting in the reduced mobility and conductivity. Hence a higher THz transmission behavior was detected. Such THz field-induced nonlinear absorption can be further scaled along with increasing the THz field strength, which is in good agreement with previous results[15].

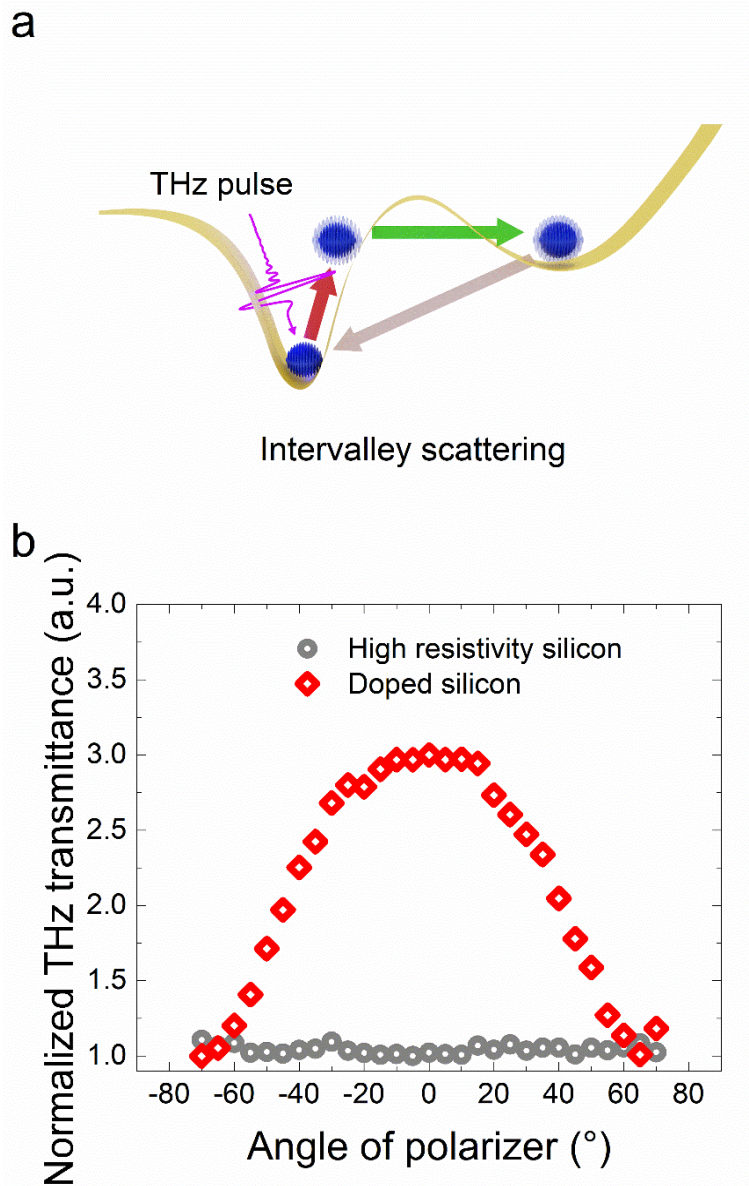

**Fig. 6 Nonlinear absorption in doped silicon sample induced by high energy THz pulses.**
**a,** Intervalley scattering microscopic schematic. **b,** Normalized THz transmittance (THz transmittance divided by the minimum transmittance) as a function of the THz polarizer angles. Obvious field sensitive nonlinear behavior can be observed.

## Conclusion and outlook

In closing, we have theoretically and experimentally investigated the realization of generating >1 mJ super high energy THz radiation based on the TPF technique in LN crystals driven by high power ultrashort femtosecond laser pulses. Through tailoring the pump laser spectrum, chirping the pump pulses, and cryogenically cooling the emission crystals to enhance the effective interaction length and reduce the THz photon absorption by lattice vibrations, we successfully achieved new record numbers of 1.4 mJ high energy THz pulse with an 800 nm-to-THz conversion efficiency of 0.7% from LN crystals. Such high energy THz pulses with ~1 ps single-cycle pulse duration can produce a free-space electric field strength of 6.3 MV/cm accompanying by an ultrashort magnetic pulse with a peak field of ~2.1 Tesla, which can offer multifaceted capabilities for unveiling extreme nonlinear THz phenomena and field-controlled light-matter interaction physics. As a preliminary start and a proof-of-principle experiment, we employed THz field-induced nonlinear absorption in doped silicon materials, and the experimental results verified the powerful capabilities of our THz source, which paves the way for more related nonlinear observations under application of such high energy THz sources.

The Ti:sapphire pump laser in this experiment can produce >500 mJ energy, but we were far from using the maximum energy for THz generation in this case due to the design limitation of aiming for ~1 mJ THz energy. However, through this experimental verification and theoretical prediction, we have the confidence to generate much higher THz energies in the near future. Furthermore, along with the rapid development of ultrafast laser technology, we believe LN-based solid-state THz sources will become more powerful and accessible, which makes us envision that more abundant extreme THz science and applications can be explored and discovered with such strong THz sources.

## Methods

**Single-shot diagnostic.**

To characterize the generated THz temporal waveform, we employed the single-shot measurement based on spectral interferometry. The setup is drowned in Fig. 2. A 1-inch sampling mirror was placed at the edge of the expanded pump beam to take portion energy as a probing beam with a spectral width of 50 nm (FWHM), which was further stretched from 40 fs to 17 ps by a grating stretcher. A delay line was used to synchronize the probing optical beam and the radiated THz signal. Finally, the probing beam was focused together with the THz beam onto a 1 mm thick electro-optic crystal ZnTe by OAP3 with a focal length of 2-inch. The Glan prism (GP1) was to purify the polarization of the probing beam. The lens L2 was to collimate the probing beam. After passing through a 4 mm thick alpha-BBO crystal, the probing beam was divided into two identical orthogonal pulses with a time delay of 1.67 ps along its fast and slow axes. Spectral-domain interference occurred at a second Glan prism (GP2) whose polarization was orthogonal to the GP1, and the spectrum was then collected and analyzed by a high-precision grating spectrometer (SpectraPro 2750, Princeton Instruments). The THz temporal transients were extracted by decoding the interference spectrum recorded by the spectrometer. More details can be found in Supplementary Information.

**THz radiation and laser spectrum variation theory.**

According to Maxwell's equations, considering that the complex envelope of the electric field of light in frequency domain changes slowly with the propagation distance inside the emission crystals, the wave equation in the frequency domain can be described as follows under the slowly varying envelope approximation,

$$\frac{\partial E_{env}(\omega, z)}{\partial z} = \frac{-j\mu_0 \omega^2}{2k(\omega)} P_{NL}(\omega, z) e^{jk(\omega)z} \qquad (7)$$

According to the nonlinear Lorentz oscillator model of the non-centrosymmetric medium, the quantitative mathematical expression of the macroscopic second-order nonlinear polarization can be derived, when the third- and higher-order nonlinear effects are neglected,

$$P_{NL}(\omega, z) = P^{(2)}(\omega, z) = \frac{Ne^3 a}{2\pi m^2 D(\omega)} \int_{-\infty}^{+\infty} \frac{E_{env}(u, z) e^{-jk(u)z} E_{env}(\omega - u, z) e^{-jk(\omega-u)z}}{D(u) D(\omega - u)} du \qquad (8)$$

Combining the frequency-domain wave equation under the slowly varying envelope approximation and the mathematical expression of the second-order nonlinear polarization, we can directly obtain the frequency-domain second-order nonlinear wave equation under the slowly varying envelope approximation,

$$\frac{\partial E_{env}(\omega,z)}{\partial z} = \frac{-j\mu_0\omega^2}{2k(\omega)} \frac{Ne^3 a}{2\pi m^2 D(\omega)} \int_{-\infty}^{+\infty} \frac{E_{env}(u,z)e^{-jk(u)z} E_{env}(\omega-u,z)e^{-jk(\omega-u)z}}{D(u)D(\omega-u)} du \cdot e^{jk(\omega)z} \quad (9)$$

Due to the little variation of the LN refractive index in the optical and THz frequency regions and the phase-matching condition almost satisfying during the laser propagation process inside the crystal through tilting pulse front technique, we can integrate the influence of the tilting angle into the wavenumber, which is described as

$$\beta(\omega-u) + \beta(u) = \beta(\omega) \quad (10)$$

Under phase-matching condition, the frequency-domain second-order nonlinear wave equation under the slowly varying envelope approximation can be further derived to the following form

$$\frac{\partial E_{env}(\omega,z)}{\partial z} = \frac{-j\mu_0\omega^2}{2k(\omega)} \frac{Ne^3 a}{2\pi m^2 D(\omega)} \int_{-\infty}^{+\infty} \frac{E_{env}(u,z)e^{-\alpha(u)z} E_{env}(\omega-u,z)e^{-\alpha(\omega-u)z}}{D(u)D(\omega-u)} du \cdot e^{\alpha(\omega)z} \quad (11)$$

where $j$ is the unit of imaginary number, $\mu_0$ denotes vacuum permeability, $k(\omega)$ is the wavenumber which can be described by $k(\omega) = \beta(\omega) - j\alpha(\omega) = \omega\sqrt{\varepsilon\mu_0}$, where $\alpha(\omega)$ presents absorption coefficient, $N$ denotes atomic number density, $e$ is the electron charge, $m$ denotes electron mass, and $a$ presents nonlinear coefficient for nonlinear Lorentz oscillator model. The factor $D(\omega)$ is described by $D(\omega) = -\omega^2 + \omega_0^2 + j\gamma\omega$, where $\omega_0$ is the eigenfrequency and $\gamma$ is damping rate. Therefore, we can quantitatively reproduce the experimental phenomena with respect to the variation of the pump laser spectrum and interpret the main mechanisms during high energy THz radiation from LNs via the TPF technique.